\documentstyle[preprint,prd,aps,floats,epsfig]{revtex}
\tighten
\begin{document} 
\draft
\preprint{
IMPERIAL/TP/97-79
}
\title{Painless causality in defect calculations}
\author{Charlotte Cheung and Jo\~ao Magueijo}
\address{The Blackett Laboratory, Imperial College,
Prince Consort Road, London SW7 2BZ, UK}
\maketitle

\begin{abstract}
Topological defects must respect causality, a statement leading to
restrictive constraints on the power spectrum of the total cosmological
perturbations they induce. Causality constraints have for long been
known to require the presence of an under-density in the surrounding 
matter compensating the defect network on large scales. This so-called
compensation can never be neglected and significantly complicates 
calculations in defect scenarios, eg. computing cosmic microwave background 
fluctuations. A quick and dirty way to implement the compensation 
are the so-called compensation fudge factors. Here we derive the 
complete photon-baryon-CDM backreaction effects in defect scenarios. 
The fudge factor comes out as an algebraic identity and so we drop
the negative qualifier ``fudge''. The compensation scale is computed
and physically interpreted. Secondary backreaction effects exist, 
and neglecting them constitutes the well-defined approximation scheme 
within which one should consider compensation factor calculations.
We quantitatively assess the accuracy of this approximation, and
conclude that the considerable pains associated with improving on it
are often a waste of effort.
\end{abstract}

\date{\today}
 
 
\renewcommand{\thefootnote}{\arabic{footnote}}
\setcounter{footnote}{0}

\section{Introduction}
When a defect network is formed, causality and energy conservation
demand that there must be a compensating under-density in the background
matter and radiation.  This compensation is exactly anti-correlated
with the defects and is of a comparable intensity. It acts as a source 
for the gravitational potential which in turn drives radiation
perturbations. For this reason the
compensation cannot be ignored in CMB calculations.  In previous work
\cite{alsteb,us} the compensation has been included by 
use of compensation fudge
factors. These ensure that the overall perturbations have a
large-scale behaviour consistent with the causality constraints
\cite{trasch,traschk4}. However their exact form was never 
justified by an analytical
identity, and hence the words ``fudge factor'' qualifying them.

In this paper we show that by choosing a suitable gauge it is possible to
derive analytically an expression for the compensation which contains
a term in the form of the previously discussed fudge factor.  Along
with the compensation factor we find that there are also
terms representing the backreaction from baryons, CDM and radiation
which cannot be predicted a priori. We find however that these terms 
are sub-dominant compared to the defects for any reasonable values 
of the Hubble constant and the
baryon fraction today.  The physical implication is that it is the
defect rather than any other component which dominates the spectrum of
radiation perturbations.  Hence by choosing a suitable
defect it is possible to have a range of spectra which do not
necessarily have the characteristic out of phase spectra associated
with defect calculations, as shown in \cite{us}.

This paper is organised as follows. In Section~\ref{sec2}
review the equations for a system of photons, baryons,
CDM, and defects, in the tight-coupled approximation. 
We present a trick for easily considering a defect component 
within a multi-fluid formalism, such as the one presented in Kodama and Sasaki
\cite{KS}. We show how the various gauge-invariant formulations
correspond to nothing but different choices of full gauge fixing.
We look into all popular choices of gauge. We argue in favour of the
flat slicing gauge for the discussion of feedback mechanisms,
the compensation, or the causality constraint. Then in Section~\ref{exact}
we present an algebraic manipulation which allows splitting 
compensation into two terms, one of which is purely a compensation
fudge factor. We evaluate then, in Section~\ref{calc}, the quantitative
impact of dropping the terms other than the compensation factor term.
We find that one would need rather extremely values of the Hubble
constant and baryon fraction for these extra terms to have much 
qualitative effect. In the concluding section we digress on the
metaphysical implications of this results, and the practical
application which we intend to give it.

\section{The basic equations}\label{sec2}
We consider the epoch before recombination when it is a good approximation
to treat all the cosmic components as a fluid. This is the so-called
tight-coupled approximation, which we shall use to develop all our
arguments. We will consider a scenario where the Universe
is made up of radiation (corresponding variables labelled by $\gamma$), 
baryons ($b$), cold dark matter ($c$), and a defect component ($s$)
where the baryons and radiation are tightly coupled (see eg.~\cite{HS} for 
a quantitative definition). 

Whenever we intend to show the generality of our results we shall also
consider a set of extra generic components, denoted by $\alpha$, which
may be neutrinos or whatever personal taste requires.
In order to extract intuition from our arguments we shall sometimes
assume that only defects and radiation are relevant, a statement
valid deep in the radiation dominated epoch. The actual calculations
will however always be valid considering the other components.
This is necessary for generality, since matter-radiation equality
may (and in fact usually does) happen before recombination.

We use the gauge-invariant formalism in all the guises discussed
in Kodama and Sasaki\cite{KS} (KS from now on). 
These different formulations, one
should point out, correspond to different choices of gauge. A set of
a priori gauge dependent variables defined in a fully fixed gauge is
of course gauge invariant. The different possible gauge-invariant 
contrast variables correspond to nothing but the density contrast
as measured in different fully fixed gauges. There is sometimes
an inappropriate feeling that ``gauge-invariant'' and ``gauge-dependent''
methods are two separate tool boxes. They are in fact one and the same
thing. The only exception to this statement is the synchronous gauge 
\cite{efst}. This choice of
gauge leaves a residual gauge freedom and therefore variables
defined in synchronous gauge can never be related algebraically to gauge
invariant variables (see \cite{Bma} for a good dictionary).

In this paper we adopt a multi-lingual attitude towards cosmic
perturbation formalisms. 
We shall use a variety of density contrast variables
$\Delta_\alpha$ for a generic component $\alpha$. These will
be indexed in the following way. No extra index after the component 
index $\alpha$ denotes density contrast in the $\alpha$-component 
rest frame (not in the total rest frame which we found a mess
in the presence of defects). 
An index $s$ denotes the density contrast in the Newtonian
slicing (where the perturbed cosmological flow appears to have no shear). 
An index $g$ denotes the density contrast in the flat slicing
(where the equal time slices have no scalar curvature). 

We consider all these different gauges in order to connect
with previous work. However the main remark in this and the next section 
is that by choosing the flat slicing contrast variables two desirable 
(at least for defect practitioners) features may be achieved.
Firstly, one gets rid of potential time derivatives
in the equations for the radiation (and also for baryons and CDM, 
but this can also be achieved in other gauges).
This is a major technical improvement on the formalism. It
allows performing all calculations invoking only defect stress energy
components and not their time derivatives. Structure functions
for time derivatives are notoriously noisier in defect
simulations.

Secondly, as shown in the next Section, in the flat slicing
gauge the radiation backreaction naturally separates into two terms. 
The first is required by the Traschen integral constraints \cite{trasch}
and cannot be set to zero in any approximation scheme, otherwise
causality is grossly violated. 
We will however find an exact expression for this term made up of
a factor (independent of the perturbation variables) times a set of 
defect variables. This factor happens to have the same form
as the ``compensation fudge factor''. Furthermore the ``compensation scale'',
left undetermined in compensation factor calculations, can be computed.
Hence fudging is an exact equality made obvious in the flat-slice
gauge.  The second backreaction term is truly unpredictable, 
but we will be able to show that it is not required by causality, 
and it is qualitatively sub-dominant. 

\subsection{A defect component in a multi-fluid formalism}
We shall add to the multi-fluid formulation of Kodama and Sasaki
a defect component. This can be best implemented by noticing that
defects have no background stress-energy.  We may then regard
defects as a fluid for which the background energy and pressure
are zero, the perturbation variables are infinite, but the product
of the background and perturbation variables is finite and equals
the defect stress-energy.

More mathematically let the defects stress-energy 
tensor $\Theta_{\mu\nu}$ be a pure scalar so that it may be written as
\begin{eqnarray}
  \Theta_{00}&=&\rho^s\nonumber \\
  \Theta_{0i}&=&k_iv^s\nonumber\\
  \Theta_{ij}&=&p^s\delta_{ij}+(k_i
                k_j-{1\over 3}\delta_{ij}k^2)\Pi^s
\end{eqnarray}
Then let us consider a component $\alpha=d$ with $\rho_d=
p_d=0$, infinite perturbation variables (eg. $\delta_d
=\infty$), but finite products of the two. From the way perturbation
variables are defined in KS \cite{KS} from the stress-energy
tensor we can then write
\begin{eqnarray}
  a^2\rho_d\delta_d&=&\rho^s\nonumber\\
  a^2(\rho_d+p_d)v_d&=&kv^s\nonumber\\
  a^2p_d\Pi^T_d&=&k^2\Pi^s\nonumber\\
  a^2p_d\Pi^L_d&=&p^s
\label{defectvar}
\end{eqnarray}
Because the background stress-energy of defects is zero, defect
variables are gauge-invariant by themselves. However care must 
be taken when identifying KS defect variables with defect
variables. For instance, it may happen that a gauge invariant 
density contrast variable is given by a combination of defect variables.
Using Eqns.~\ref{defectvar} we can find the identifications:
\begin{eqnarray}
  a^2\rho_d\Delta_d&=&\rho^s+3hv^s\\
  a^2\rho_d\Delta_{sd}&=&a^2\rho_d\Delta_{gd}=\rho^s
\end{eqnarray}
For all other variables there is no ambiguity, as the extra
terms required to turn gauge-dependent variables into gauge
independent ones simply vanish. For instance
\begin{eqnarray}
  a^2(\rho_d +p_d)V_d&=& a^2(\rho_d +p_d)(v_d-\dot H_T/k)=kv^s\\
  a^2p_d\Pi^L_d&=&p^s\\
  a^2p_d\Pi^T_d&=&k^2\Pi^s
\end{eqnarray}
The conservation equations for the defect component may be written as;
\begin{eqnarray}
{\dot\rho}^s+h(3p^s+\rho^s)+k^2v^s&=&0\label{cons1}\\
{\dot v}^s+2hv^s-p^s+{2\over 3}k^2\Pi^s&=&0\label{cons2}\; .
\end{eqnarray}
which can be derived from the conservation equations in KS
with the identifications made above. 
The gauge-invariant potentials $\Phi$ and $\Psi$ can be obtained
from  the Einstein's equations in KS. These are now sourced 
by a total density contrast and anisotropic stress which contains
defects. We choose to separate the defects from
all other components. Hence in all formulae in KS containing
totals the following replacements should
be introduced
\begin{eqnarray}
  a^2\rho\Delta_T&\rightarrow
    &a^2\rho\Delta_T+\rho^s+3hv^s=a^2\sum\rho_\alpha\Delta_\alpha
    +\rho^s+3hv^s\\
  a^2(p+\rho)V_T&\rightarrow
    &a^2(p+\rho)V_T+kv^s=a^2\sum(p_\alpha+\rho_\alpha)V_\alpha
    +kv^s\\
  a^2p\Pi^T_T&\rightarrow
    &a^2p\Pi^T_T+k^2\Pi^s=a^2\sum p_\alpha\Pi^T_\alpha + k^2\Pi^s
\end{eqnarray}
Bearing this in mind, the Einstein equations in the presence of
defects may now be read off from KS as
\begin{eqnarray}\label{poteq}
  k^2\Phi&=&4\pi{\left(a^2\rho\Delta_T +\rho^s+3
            h v^s\right)}\label{poteq1}\\
  \Phi+\Psi&=&-8\pi{\left(a^2{p\Pi^T_T\over k^2}+\Pi^s
               \right)}\label{poteq2}
\end{eqnarray}
In the scenario we are considering the total density contrast,
putting defects aside, is given by
\begin{equation}
  \rho\Delta_T=\rho_b\Delta_b+\rho_{\gamma}\Delta_{\gamma}+\rho_c\Delta_c
\end{equation}
The fluids viscosity $\Pi^T_T$ is entirely due to the photons brightness
quadrupole \cite{HS}:
\begin{equation}
  \Pi={12\over5}\Theta_2
\end{equation}
and can be set to zero in the tight-coupling limit. 

\subsection{The Newtonian slicing equations}
The Newtonian slicing equations for the radiation are what
leads to the Hu and Sugyama (HS) formalism \cite{HS}.
During tight-coupling the photon system is described in HS by the
monopole and dipole components of the brightness function, $\Theta_0$
and $\Theta_1$. In the fluid description this corresponds to the
Newtonian slicing variables 
\begin{eqnarray}
  \Theta_0&=&\Delta_{s\gamma}/4\\
  \Theta_1&=&V_\gamma
\end{eqnarray}
It can be checked that, with this identification, the conservation
equations in KS for radiation become the HS equations:
\begin{eqnarray}\label{tighteqns}
  {\dot\Theta_0}&=&-{k\over 3}\Theta_1-{\dot \Phi}\nonumber\\
  {\dot\Theta_1}&=&-{\dot R \over 1+R}\Theta_1+{k\over 1+R}\Theta_0
                   +k\Psi
\end{eqnarray}
where $R={3\over 4}{\rho_b\over\rho_{\gamma}}$ is the
scale factor normalised to $3/4$ at photon-baryon equality.
It is for these equations that HS propose a WKB solution,
which was used in the study of defect Doppler peaks in \cite{us}.
If one uses this gauge for the radiation one must however
change to the comoving gauge before computing the potentials
$\Psi$ and $\Phi$. This can be easily done by means of
\begin{equation}
  \Delta_{\gamma}=4{\left(\Theta_0+h{\Theta_1\over k}\right)}
\end{equation}

\subsection{The comoving gauge}
In Hu and White (HW) \cite{hw} the issue of backreaction is addressed
in the comoving gauge (the word gauge being replaced by ``representation''). 
A temperature variable is defined such that
\begin{equation}
 {\cal T}=\Delta_\gamma/4=\Theta_0+h{\Theta_1\over k}
\end{equation}
and an horrible set of equations is derived for them.
The comoving gauge has the advantage that it is the natural gauge
for representing the baryons, since in tight coupling baryons
and photons share the same rest frame. As shown in HS and HW,
the baryons' density contrast $\Delta_b$ and velocity
$V_b$ satisfy the conditions
\begin{eqnarray}\label{baryons}
  {\dot \Delta_b}&=&{3\over 4}{\dot \Delta _{\gamma}}\label{adiab}\\
  V_b&=&V_{\gamma}
\end{eqnarray}
which can be rewritten by defining the entropy as $s=\Delta_b-(3/4)
\Delta_{\gamma}$, and rewriting the first equation as $\dot s=0$.

The comoving gauge is also the gauge where the Traschen integral
constraints are written \cite{trasch,traschk4}.
In an expanding Universe a set of energy conservation
laws apply to perturbation variables \cite{trasch}. When applied 
to a Universe which is initially unperturbed, and 
then causally made inhomogeneous, these constraints translate 
into a stringent requirement on the large scale power 
spectrum of these perturbations \cite{traschk4}. This requirement
is roughly that the power spectrum of the total energy 
perturbation in the comoving
gauge goes like $k^4$ for small $k$. In the presence of defects
the energy density subject to this law is:
\begin{equation}
  {\cal U}=a^2\rho\Delta_T+\rho^s+3hv^s
\end{equation}
which is also the source of the gauge-invariant potential $\Phi$.
Hence one can rephrase the causal constraint as the requirement
that $\Phi$ goes to a constant at low $k$ (white-noise).

The lack of superhorizon correlations in the defect network requires
the power spectrum $P(\rho^s)$ to have a white noise low $k$ tail.
Energy conservation Eqn.~(\ref{cons1}) requires that $v^s$ also
have a white noise low $k$ tail. Hence the causal constraint
entails the need for the compensation: a low $k$ white-noise tail 
in the  power spectrum of non-defect matter. The compensation must be
exactly anticorrelated with the defects' tail. The quantity to be cancelled is
$\rho^s+3hv^s$, and not just $\rho^s$. The density forced
to have a $k^4$ power spectrum is  ${\cal U}$
and not just a combination of $\Delta_T$ and $\rho^s$.

\subsection{The flat-slicing gauge}
We can also define a temperature perturbation in the flat-slicing
gauge \cite{note}:
\begin{equation}
  \Delta_0=\Theta_0+\Phi=\Delta_{g\gamma}/4
\end{equation}
In this gauge the photon equations are:
\begin{eqnarray}\label{tighteqnsft}
  {\dot\Delta_0}&=&-{k\over 3}\Theta_1\nonumber\\
  {\dot\Theta_1}&=&-{\dot R \over 1+R}\Theta_1+{k\over 1+R}(\Delta_0-\Phi)
                   +k\Psi
\end{eqnarray}
As announced before one gets rid of the potential time derivatives 
in this gauge. We will also find it useful to represent
CDM in this gauge, so that CDM equations are:
\begin{eqnarray}\label{cdmft}
  \dot\Delta_{gc}&=&-kV_c \nonumber\\
  \dot V_c&=&-hV_c+k\Psi
\end{eqnarray}
There is a good mathematical reason why this gauge may be better
for discussing causality and compensation issues. This is the gauge
where the fluid equations of motion more resemble Minkowski space-time
equations of motion. Hence energy variables in this gauge are akin
to the pseudo-energy usually defined in the synchronous gauge,
and used to introduce the compensation \cite{PST,james}.

\section{Exact compensation factors}\label{exact}
We start with an algebraic remark.
The source for the Einsteins equations are energy density variables
in the comoving gauge. In the scenario we are considering
\begin{eqnarray}
  k^2\Phi&=&4\pi{\{a^2\rho(\Omega_b\Delta_b+\Omega_\gamma\Delta_\gamma
            +\Omega_c\Delta_c+\Omega_\alpha\Delta_\alpha)+\rho^s+3hv^s\}}\\
  \Phi+\Psi&=&-8\pi\Pi^s
\end{eqnarray}
where $\alpha$ represents any other component we may have forgotten,
with a sum over $\alpha$ implied, if need be.
Now let us express baryons in terms of photons by means of
Eq.~(\ref{adiab}) written as
\begin{equation}
  \Delta_b={3\over 4}\Delta_\gamma +s
\end{equation}
and write all other sources in the flat-slice gauge:
\begin{eqnarray}
  \Delta_\gamma&=&4{\left(\Delta_0-\Phi+h{\Theta_1\over k}\right)}\\
  \Delta_c&=&\Delta_{gc}+3{\left(h{V_c\over k}-\Phi\right)}\\
  \Delta_\alpha&=&\Delta_{g\alpha}+3(1+w_\alpha){\left(h{V_\alpha\over k}
                -\Phi\right)}
\end{eqnarray}
With these rearrangements, the first Einstein equation becomes
\begin{eqnarray}
  k^2\Phi=4\pi{\Big(}a^2\rho{\Big(}4\Omega_{\gamma}(1+R){\left(\Delta_0
          -\Phi+h{\Theta_1\over k}\right)}+\Omega_bs+&&\nonumber\\
          \Omega_c{\left(\Delta_{cg}-3\Phi+3{V_c\over k}\right)}
          +\Omega_\alpha{\left(\Delta_{\alpha g}+3(1+w_\alpha)
              {\Big (}-\Phi+{V_\alpha\over k}{\Big)}
            \right)}{\Big)}+\rho^s+3hv^s{\Big )}&&
\end{eqnarray}
The source term can now be split into 3 components:
\begin{equation}
  k^2\Phi=S+S_1+S_2
\label{aux}
\end{equation}
where
\begin{eqnarray}
  S&=&4\pi(a^2\rho_bs+\rho^s+3hv^s)\\
  S_1&=&-4\pi a^2\rho{\left( \Omega_\gamma(1+R)+3\Omega_c+3(1+w_\alpha)
        \Omega_\alpha\right)}\Phi\\
  S_2&=&4\pi a^2\rho(4\Omega_\gamma(1+R)(\Delta_0+h\Theta_1/k)
        +\Omega_c(\Delta_{gc}+3hV_c/k)+
        \Omega_\alpha(\Delta_{g\alpha}+3hV_\alpha/k)
\end{eqnarray}
$S$ is made up of sources which drive the radiation-baryon-CDM
system but which are external to them. We call it the external source.
This may be a topological defect. An entropy perturbation 
may also be regarded as external since it evolves independently
of all other perturbations (according to $\dot s=0$). There is a fundamental
difference between defects and entropy perturbations. Entropy
perturbations satisfy $\dot s=0$. Defect sources, on the contrary, satisfy
Eqns.~(\ref{cons1}) and (\ref{cons2}). 

The photon-baryon-CDM system is also driven by a backreacting 
term, here split as $S_1+S_2$. This reflects the fact that baryons,
photons and CDM are driven by a potential which they are a source of.
There is therefore a (linear) feedback effect which jeopardises
for instance the use of the WKB solution in HS for defects.
One could hope that defects are the main driving force, and try
to neglect backreaction. However this back-reacting term incorporates
the compensation. Setting $S_1+S_2$ to zero is therefore an approximation
which can never make sense, as it would imply a gross violation
of the causality constraint. The potential $\Phi$ power spectrum
would diverge like $1/k^4$ at small $k$ rather than go to white noise.

However, in the flat-slice gauge we have an algebraic bootstrap
which one may hope already reflects most of the physical feedback mechanism.
This bootstrap is created by the term $S_1$. Let us first
set $S_2=0$. Then all the backreaction is predictable
and fully determined by the defect sources. $S_1$ may be passed to
the left hand side of the Einstein equation (\ref{aux}) and 
be incorporated in an
equation where the external sources are simply multiplied by a factor
independent of photon, baryon or CDM variables. More precisely
\begin{equation}\label{fudgepot}
  k^2\Phi=4\pi\gamma_c(a^2\rho_bs+\rho^s+3hv^s)
\label{fudgephi}
\end{equation}
where 
\begin{eqnarray}
  \gamma_c&=&{1\over 1+(\chi_c/x)^2}\\
  \chi_c^2&=&{3\over 2}(h\eta)^2(4\Omega_{\gamma}(1+R)+3\Omega_c
              +3(1+w_\alpha)\Omega_\alpha)
\end{eqnarray}
The backreaction encoded in $S_1$ is not an independent feedback
mechanism operating in the fluid and imprinting a fixed signature
in the photons' power spectrum for any defect theory. 
From this term we can never expect to
derive an out-of-phase signature as the one attributed to defects
in \cite{hw}. The backreaction contained in this term is fully 
driven by the defects alone, and can be made to behave in whatever 
way we want by properly designing the defect. It is not surprising
that by considering only this backreaction effect we can place the primary
Doppler peak anywhere, including the adiabatic and out-of-phase 
positions \cite{us}. 
On the other hand by considering this term one is already taking into
account the causality constraint. The potential $\Phi$ according
to the new equation (\ref{fudgephi}) already goes to white noise
at small $k$.

If $S_2\neq 0$ then one must add to equation (\ref{fudgepot})
an extra source term, so that
\begin{equation}\label{fudgepottot}
  k^2\Phi=\gamma_c(S+S_2)
\end{equation}
This is an exact expression. The compensation factor has appeared as
a result of algebra and the compensation scale is a well defined
quantity dependent only on the unperturbed cosmological expansion dynamics. 
The compensation factor approximation is now 
the claim that we can set to zero $S_2$. This is a well defined 
statement which we should be able to assess quantitatively. 
The term in $S_2$ is the truly unpredictable backreaction. 
The $S_1+S_2$ split has allowed us to separate what is truly
a problem and what is not. By doing so we have implemented an 
approximation scheme
($S_2\approx 0$) where we may avoid the feedback problem
without immediately doing something stupid, like violating the causality
constraint.

Physically what this algebraic manipulation amounts to is the
realization, made obvious in the flat-slice gauge, that the
compensation is made up of 3 terms. One is the energy perturbation
as it appears in a gauge where equal time surfaces appear to have
no curvature. The other two are a potential perturbation describing
the curvature of these slices, and a velocity term. The potential
term is caused by the defects as well, and by dropping all other
contributions we obtain a non pathological approximation scheme
where the compensation is fully gravitational and perfectly
correlated to the defect network.

The compensation scale $\chi_c$ varies from ${\sqrt{6}}$ in the
radiation epoch to ${\sqrt{18}}$  in the matter epoch.
The compensation scale depends purely on the expansion
kinematics and is affected by the matter radiation transition. 
It can be written as
\begin{equation}
  \chi_c^2=3\eta^2(h^2-\dot h)=12\pi a^2(p+\rho)\eta^2
\end{equation}
and if $h=\alpha/\eta$ then $\chi_c^2=3\alpha(\alpha+1)$.

\section{Quantitative argument in favour of compensation factors}\label{calc}
The question remains of how good an approximation setting $S_2=0$
is. We address this question quantitatively. For definiteness
we use the source defined in \cite{hsw}:
\begin{eqnarray}
  p^s&=&{1\over \eta^{1/2}}{\sin{Ak\eta}\over Ak\eta}\\
  \Pi^s&=&0
\end{eqnarray}
bearing in mind that this ansatz may preclude arbitrary shifts
in peaks' positions. This property is far from general, 
as shown in \cite{neil1,neil2}. This issue is beyond the scope 
of this paper, but will be addressed in a future 
publication~\cite{confusion}. We then solve equations (\ref{tighteqnsft})
for radiation, (\ref{cdmft}) for CDM, (\ref{cons1}) and (\ref{cons2})
for the remaining defect variables, and (\ref{poteq1}) and (\ref{poteq2})
for the potentials (with (\ref{poteq1}) rewritten as in (\ref{fudgepottot})).
The baryons are solved implicitly by the tight-coupled conditions
(\ref{baryons}). By including  $S_2$ as a source of the potential
$\Phi$ in (\ref{fudgepottot}) one is considering the full backreaction
effects, due to baryons, radiation, and CDM. One thus obtains the
exact solution to this problem. By dropping the term in $\Delta_{gc}$
and $V_c$ in $S_2$ one neglects the effects of CDM fluctuations on the CMB
fluctuations. By setting $S_2=0$ one neglects the effects of backreaction
altogether. In the last approximation the source is truly external,
and is compensated purely by defect gravitational effects. This is the 
compensation factor approximation. 

We have solved this problem for various values of the Hubble
constant and baryon content of the Universe. These are parameterised
by $h$, so that the Hubble constant nowadays is 
$H_0=100h {\rm Km s}^{-1} {\rm Mpc}^{-1}$,  and 
$\Omega_b=\rho_{b0}/\rho_0$, the baryons density fraction nowadays. 
We expressed our results in terms of
the effective temperature $\Theta_0+\Psi=\Delta_0+(\Psi-\Phi)$
and the dipole $\Theta_1$ at last scattering $\eta=\eta_*$. 
This is because these are the quantities which are then projected 
onto $C_l$'s by free-streaming after the last scattering
surface \cite{HS}. We have solved the full problem and compared the full answer
with the effect of dropping CDM, and dropping all CDM-photon-baryon
backreaction. In all plots these 3 calculation schemes will be
represented by dash, dots, and lines, respectively. 

\begin{figure}\label{fig1}
\begin{center}
    \leavevmode
        {\vbox %
{{\epsfxsize = 15cm\epsfysize=15cm
    \epsffile {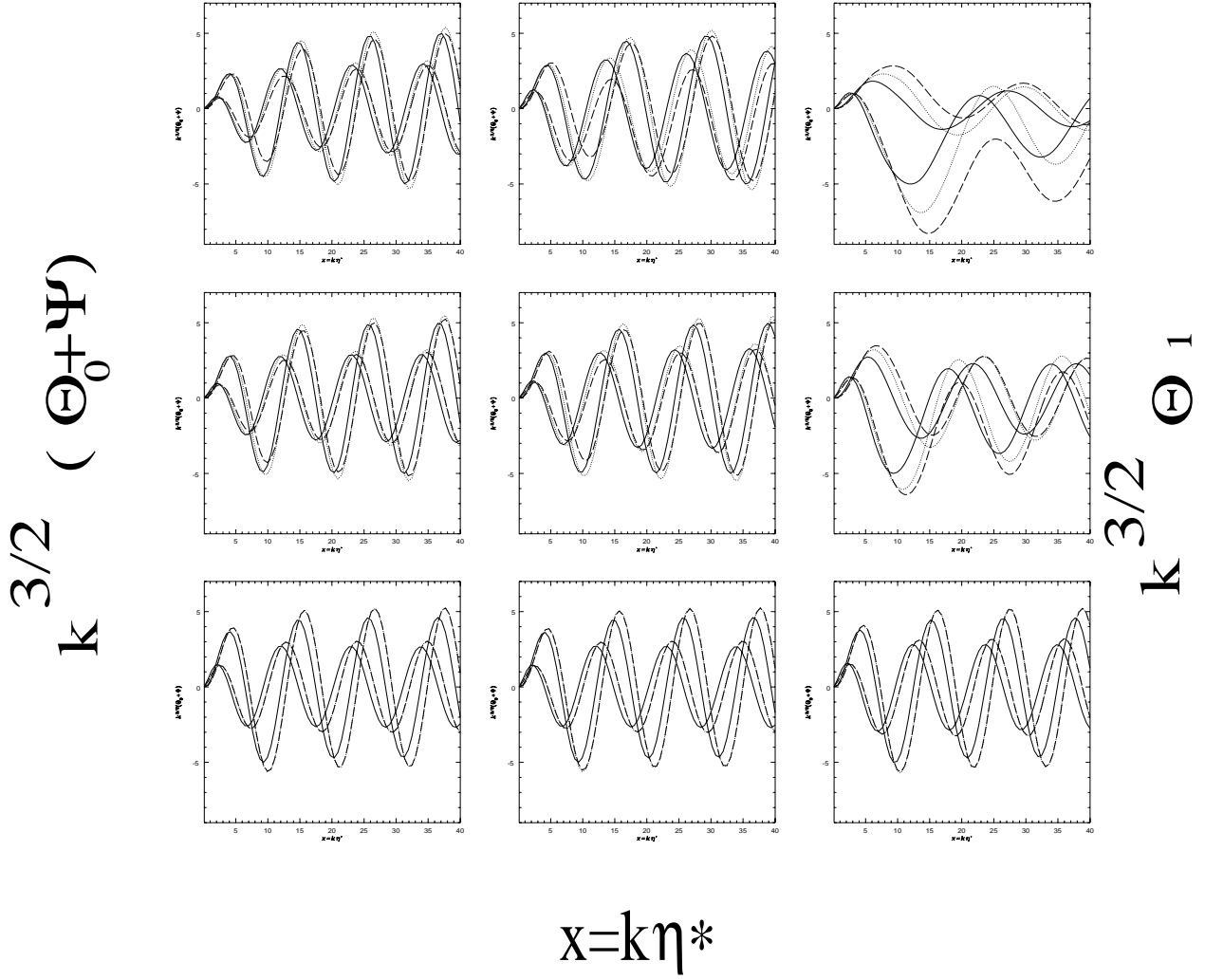} }}}
\end{center}
\caption{We plot $k^{3/2}(\Theta_0+\Psi)$
(curves which go negative first) and $k^{3/2}\Theta_1$ with the
following conventions: 
the compensation factor approximation in lines, photon-baryon 
backreaction included in dots, all backreaction included
in dash. From left to 
right $\Omega_b=0.005,0.05,0.5$. 
From bottom to top $h=0.1,0.5,0.8$.}
\end{figure}

The results for a source with $A=1$ are plotted in Figure~\ref{fig1},
and we now comment on them.
We have chosen extreme values for $\Omega_b$ and $h$ in order to 
emphasise the point we wish to make. For popular values of
$\Omega_b$ and $h$ the compensation factor approximation works
very well. Clearly one needs rather high values for both $h$ and 
$\Omega_b$ for the compensation factor approximation to become gross. 

As $h$ increases the time between equality and last scattering increases.
As a result CDM fluctuations have time to start to grow while they still
can interfere with the tightly coupled radiation. As a result the CDM
contribution  cannot really be neglected in scenarios with a large
Hubble constant. 

As $\Omega_b$ increases the so-called acoustic signature may be imparted
on the peaks. This is an asymmetry in amplitude between odd and even
peaks resulting from baryons shifting the zero level of the oscillations.
After squaring, the peaks will appear alternately big and small.
It is interesting to note two things. First the acoustic signature appears
already in the compensation factor approximation, although less pronounced.
Secondly, if one is to pay attention to detail, then CDM is as important 
as the baryons in imprinting the full acoustic signature.

Another feature present in the spectra is the shoulder preceding the
peaks, ubiquitous in defect scenario. This is not present in $\Delta_0$
and is due to the gravitational redshift term $\Psi-\Phi$.
At large scales $\Delta_0\approx 0$ and then goes negative. The
potential term $\Psi-\Phi$ is white noise and positive as $k\rightarrow 0$,
then goes to zero. This induces a pre-peak which is not acoustic, 
but merely a gravitational redshift effect at last scattering.

All in all backreaction seems to be negligible in the qualitative
study of defects. Historically it has been known that backreaction
in defect theories can never be neglected, as a result of the causality 
constraints. However we have now shown that backreaction can be exactly split
into two terms. One is purely gravitational, makes sure that
the causality constraints are satisfied,  and can be predicted
a priori from the defects by means of a compensation factor. 
The other cannot be predicted a priori but it is not required by causality.
It merely reflects the baryons, photons,
and CDM trying to make a nuisance of themselves, rebelling against
the driving force of the defects, trying to imprint a signature
they are allowed to imprint whenever no driving force is present. 
However for all reasonable values of $\Omega_b$ and $h$ this
secondary backreaction is quantitatively sub-dominant. When driven
by defects the radiation/baryon/CDM feedback effect is normally
weaker than the external  force they are subject to.
Therefore, it turns out that in defect scenarios, once compensation 
factors are taken into account, backreaction is precisely something 
which can be neglected, certainly in any qualitative discussion of 
defect perturbations.
 
\section{Waste not - a defect factory}

In this paper we stressed the technical difficulties induced
when considering compensation in defect calculations but have shown how they
can be effectively bypassed by means of compensation factors.
The compensation results from a feedback mechanism which operates 
in the photon, CDM, baryon system when an external driving force
is applied to them.  Some of this backreaction is 
unpredictable in the sense that by specifying the driving
force one does not specify the backreaction before the system of
differential equations describing the whole system is evolved.
However the dominant compensation term is always predictable in the 
sense that all one needs to do is to multiply some appropriate
combination of defect stress energy components by a compensation
factor. 

The practical implication of this result is of course not
to simplify the numerics of evolving the tight coupled equations; this
problem is straightforward enough for simplification to be necessary. 
However there is another, more metaphysical, side to compensation
considerations in the literature: this is the belief
that backreaction effects force the Doppler peaks to be out of phase.
This could only be a theorem if the compensation was the result
of a self-regulatory feedback mechanism operating in the fluid,
regardless of the details of the force driving it. The compensation
described by the compensation factor, which we showed to be dominant,
is precisely the opposite of this. It represents a feedback effect
which is directly connected to the driving force, and which has a
scale which can be manipulated by tuning the scale of the driving
force. It was in fact shown in \cite{us} that in the compensation
factor approximation, a causal source could be designed so
as to shift the Doppler peaks about, placing them on the adiabatic 
position or anywhere to its right. 

The fact that the compensation factor approximation also seems
to reproduce other effects, such as the acoustic signature, then
provides us with the  practical application of this work. If the
compensation factor approximation works one can write down a
solution expressing the monopole and dipole in the photons at 
last scattering as a function of a defect structure function.
We can then invert this expression and instead write down a
defect structure function which could give us a given monopole
and dipole at last scattering. This trick can then be converted into 
a defect factory, allowing intelligent guesses of causal sources
which exhibit effects under study. In particular one may use
this defect factory to produce confusing defects: causal sources
exhibiting allegedly inflationary signatures, such as the 
acoustic signature. Naturally once an educated guess has
been made, one should then go and evolve the full system of 
tightly coupled equations, or even better solve a Boltzmann code 
for the proposed source. However there is no reason why one should 
go the complicated way when guessing sources. In a future publication
we shall use this approach in the study of confusing defects \cite{confusion}.
\section*{Acknowledgements}
We  would like to thank A.Albrecht and W.Hu for discussion and advice.
We also thank MRAO for use of computer facilities.
C.Cheung acknowledges support from PPARC, and J.Magueijo from
the Royal Society.


\begin{thebibliography}{99}
\bibitem{note}
Energy and temperature perturbations for radiation are connected
by a factor of 4 in any gauge simply as a result of the Stephan-Boltzmann
equation $\rho_\gamma\propto T^4$, or $
\delta\rho_\gamma/\rho_\gamma= 4 \delta T/T$.
\end{thebibliography}
\end{document}